\documentclass[prl,a4paper,groupedaddress,twocolumn]{revtex4}
\usepackage{graphicx}
\usepackage{tabularx}
\usepackage{amsmath}
\usepackage{amsfonts}
\usepackage{amssymb}

\begin{document}
\title{Surface wave generation and propagation on metallic
subwavelength structures measured by far-field interferometry}
\date{\today}
\author{G. Gay}\author{O. Alloschery}\author{B. Viaris de Lesegno}\altaffiliation[Present address:]{ Laboratoire Aim{\'e} Cotton, Campus
d'Orsay, 91405 Orsay, France}
\author{J. Weiner}
\affiliation{IRSAMC/LCAR\\ Universit\'e Paul Sabatier, 118 route
de Narbonne,\\31062 Toulouse, France} \author{H. J. Lezec}
\affiliation{Thomas J. Watson Laboratories of Applied Physics,
California Institute of Technology, Pasadena, California 91125
USA}\affiliation{Centre National de la Recherche Scientifique, 3,
rue Michel-Ange, 75794 Paris cedex 16, France}
\begin{abstract}
Transmission spectra of metallic films or membranes perforated by
arrays of subwavelength slits or holes have been widely
interpreted as resonance absorption by surface plasmon polaritons
(SPPs).  Alternative interpretations involving evanescent waves
diffracted on the surface have also been proposed. These two
approaches lead to divergent predictions for some surface wave
properties. Using far-field interferometry, we have carried out a
series of measurements on elementary one-dimensional (1-D)
subwavelength structures with the aim of testing key properties of
the surface waves and comparing them to predictions of these two
points of view.
\end{abstract}
\maketitle

Early reports of transmission through arrays of subwavelength
holes in thin films and membranes\,\cite{ELG98,TPL01,GTG98},
enhanced well beyond conventional expectation\,\cite{B54} have
motivated numerous attempts to explain the underlying physics of
these surprising results. Since the early experiments were carried
out on metal films, surface plasmon polariton
resonances\,\cite{Raether88,BDE03} were invoked to explain the
anomalously high transmission and to suggest new types of photonic
devices\,\cite{BDE03}. Other interpretations based on ``dynamical
diffraction" in periodic slit and hole arrays\,\cite{T99,T02} or
various kinds of resonant cavity modes in 1-D slits and slit
arrays\,\cite{CL02,VLE03} have also been proposed.  Reassessment
of the early claims by new numerical studies\,\cite{CGS05} and new
measurements\,\cite{LT04} have prompted a sharp downward revision
of the enhanced transmission factor from $\simeq 1000$ to $\simeq
10$ and have motivated the development of an alternative approach
based on a composite diffractive evanescent wave
(CDEW)\,\cite{LT04}. This model constructs a composite surface
wave from the distribution of diffracted evanescent modes (the
inhomogeneous modes of the ``angular spectrum representation" of
wave fields\,\cite{MW95}) originating at an abrupt surface
discontinuity such as a subwavelength-sized hole, slit, or groove
when irradiated by a train of plane waves. The CDEW model exhibits
three specific properties. First, the surface wave is considered
to be a composite of modes labelled by the propagation vector
component parallel to the surface, and evanescent in the direction
normal to the surface. This composite ``wave packet" exhibits
well-defined, regular nodal positions spaced by a characteristic
wavelength, $\lambda_{\mathrm{surf}}$; second, the appearance of
the first node at a distance of $\lambda_{\mathrm{surf}}/2$ from
the structured edge implies an effective phase delay of $\pi/2$
with respect to the E-field of the external driving source; and
third, an amplitude decreasing inversely with distance from the
launch site with an overall effective range of a few microns.

\begin{figure}
\centering
\includegraphics[width=0.85\columnwidth]{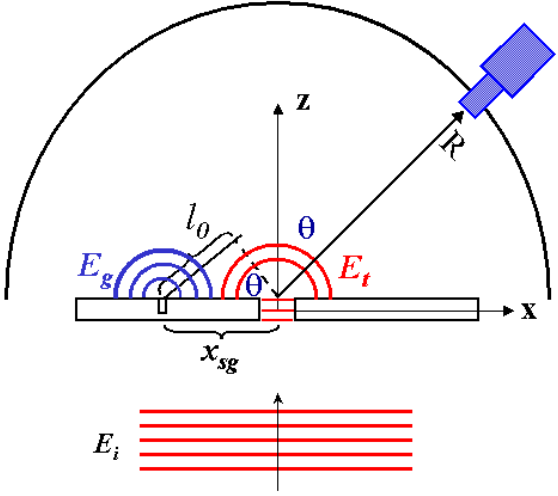}\caption{Diagram showing interfering wavefronts, optical path difference between
$E_t$ and $E_g$, and far-field
detection.}\label{Fig:InterferenceDiag}
\end{figure}

We have fabricated 1-D structures (slits and grooves with
subwavelength widths) in thin silver films deposited on fused
silica substrates designed to test these features of the CDEW
model. The optical response can be studied with the structures
facing toward (input-side experiments) or away from (output-side
experiments) a distant coherent plane-wave source. Results from
the input-side experiments, exhibiting light transmission
interference as a function of slit-groove distance, have been
reported elsewhere\,\cite{GAV06}. We report here measurements of
output-side, far-field intensity fringes arising from interference
between propagating waves transmitted through the slit and surface
waves launched at the slit but, after travelling along the
surface, reconverted to outgoing waves at the groove (see
Fig.\,\ref{Fig:InterferenceDiag}). Studies of fringe frequency,
phase, and contrast as a function of slit-groove distance and
groove depth provide new and complementary information to the
previously reported input-side experiments.

 The subwavelength structures are fabricated by focused ion beam
 (FIB) milling (FEI Nova-600 Dual-Beam system, Ga$^+$ ions,
 30keV) in a 400 nm thick layer of silver evaporated onto flat fused
 silica microscope slides.  A low beam current (50\,pA) is used in
 order to achieve surface features defined with a lateral precision
 on the order of 10\,nm and characterised by near-vertical
 sidewalls and a minimal amount of edge rounding.  Since it enables
 delivery of a variable ion dose to each pixel of the writing
 field, FIB milling conveniently allows the multiple-depth
 topography characteristic of the present devices to be formed in a
 single, self-aligned step. A 2-D matrix of structures is milled
 into the silver layer. Each matrix consists of 63 structures, nine
 columns, separated by 1.5 mm, and seven rows, separated by 2 mm.
 The first column contains only slits with no flanking grooves.
 Light transmission through the slits in this column is used to
 normalise the transmission in the remaining columns. Variations in
 transmission through each of the elements in the ``slits only"
 column provide a measure of the uniformity of the FIB fabrication
 process.  The remaining structures consist of slits flanked on one side by a groove.
 The groove-slit distance is systematically increased from the initial to final matrix positions.
 The square microscope slides themselves are 25 mm on a side and 1 mm thick.

\begin{figure}\centering
 \includegraphics[width=\columnwidth]{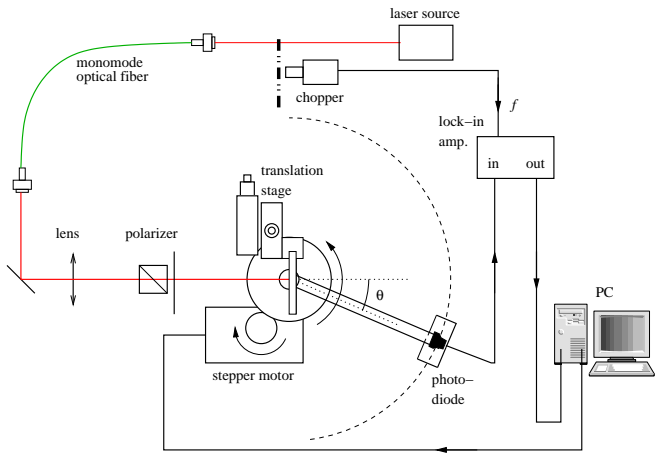}
 \caption{Goniometer setup for measuring far-field light intensity
 and angular distributions.  See text for details.}\label{Fig:goniometre}
 \end{figure}

 Measurements were carried out using a home-built goniometer shown
 in Fig.\,\ref{Fig:goniometre}.  Output from a diode laser source,
 temperature stabilised and frequency-locked to the Cs
 $^2\mathrm{S}_{1/2}(\mathrm{F}=4)\rightarrow$ $
 ^2$P$_{3/2}(\mathrm{F}=4,5)$ crossover feature in a saturated
 absorption cell, is modulated at 850 Hz by a mechanical chopper,
 injected into a monomode optical fibre, focused and linearly
 polarised (TM polarisation, H-field transverse to the slit) before impinging perpendicularly on the structure matrix mounted
 in a sample holder.  The beam waist diameter and confocal
 parameter of the illuminating source are 300 $\mu$m and 33 cm,
 respectively. Throughout this series of measurements the laser
 power density was maintained $\sim 1$W\,cm$^{-2}$. The sample holder
 itself is fixed to a precision x-y translator, and individual slit-groove structures of the 2-D matrix are
 successively positioned at the laser beam waist.  A photodiode
 detector is mounted at the end of a 200 mm rigid arm that rotates
 about an axis passing through the centre of the sample holder.  A
 stepper motor drives the arm at calibrated angular increments of
 2.05 mrad per step, and the overall angular resolution of the
 goniometer is $\simeq 4$ mrad.  The photodetector output current
 passes to a lock-in amplifier referenced to the optical chopper
 wheel. Data are collected on a personal computer that also
 controls the goniometer drive.

The structures consist of a single subwavelength slit, 100\,nm
wide flanked by one subwavelength groove. Both groove and slit are
20\,$\mu$m long. We have carried out measurements with grooves of
two different widths (100 nm and 415 nm) and depths varying from
32 nm to 256 nm. Here we report results only for the 100 nm
grooves since the wider structures yield similar results.

\begin{figure}
\centering
\includegraphics[width=0.9\columnwidth]{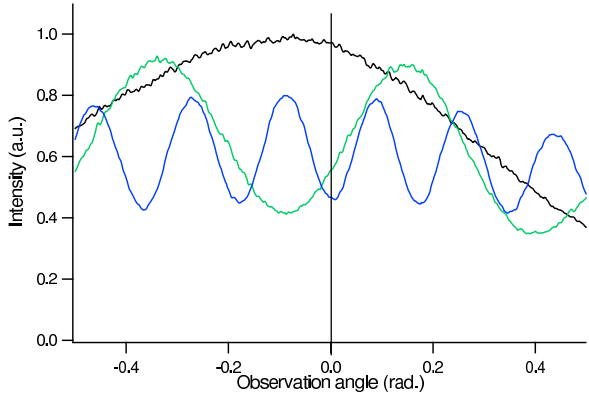}\caption{Representative output-side interference fringes for structures with 100
\,nm wide grooves at three different distances $x_{sg}$: black,
0.543 $\mu$m; green, 1.845 $\mu$m; blue, 4.991 $\mu$m.
 The phase value at zero observation angle ($\theta=0$) is used to plot $\cos\varphi$ in Fig.\,\ref{Fig:1stgOutputSidePhase}.}\label{Fig:1s1gOutputSideResults}
\end{figure}

\begin{figure}\centering
\includegraphics[width=0.85\columnwidth]{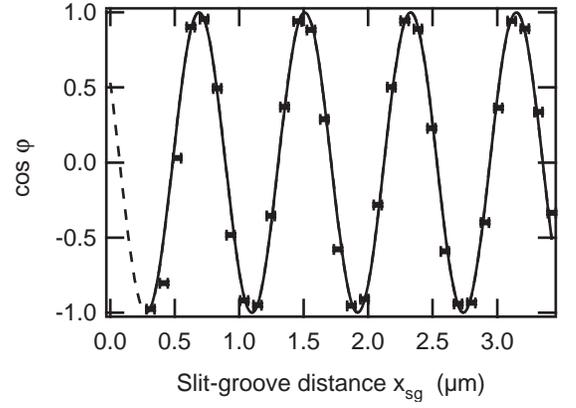}\caption{Plot of the phase $k_{\mathrm{surf}}x_{sg}+\varphi_{\mathrm{int}}$ as a function of
$x_{sg}$ with the detector oriented perpendicular to the structure
plane.  Spectral analysis of the frequency spectrum of the fringes
yields a determination of the surface wavelength
$\lambda_{\mathrm{surf}}=811\pm 8$\,nm.  Extrapolation of the
phase as $x_{sg}$ approaches zero, yields
$\varphi_{\mathrm{int}}=0.32\,\pi \pm 0.02\,\pi$.
}\label{Fig:1stgOutputSidePhase}
\end{figure}

As illustrated in Fig.\,\ref{Fig:InterferenceDiag}, the far-field
intensity pattern should exhibit interference fringes between
$E_t$ directly propagating through the slit and $E_g$ radiating
from the grooves after having been transmitted by the surface
waves $E_{\mathrm{surf}}$ launched at the output side of the slit.
The frequency and phase of the interference pattern is a function
of the optical path difference and any ``intrinsic" phase shift
(e.g. due to groove shape or surface wave phase lag) of $E_g$ with
respect to $E_t$. Figure \ref{Fig:1s1gOutputSideResults} shows
interference fringes at three representative slit-groove distances
$x_{sg}$ as a function of the goniometer detector angle $\theta$.
The fractional surface wave amplitude $\alpha$, normalised to the
incoming plane wave amplitude $E_i$,
$\alpha=E_{\mathrm{surf}}/E_i$, is estimated from the Kowarz model
\cite{K95} to be about 95\%. The remaining 5\% constitutes the
amplitude fraction $\delta$ of the light directly transmitted
through the slit $E_t$. A further fraction $\beta$ is reconverted
to propagating light $E_g$ at the groove site $x_{sg}$ and
interferes with $E_t$. The intensity, $I_{g}$, of the superposed
wavefronts can be expressed as
\begin{equation}
I_{g}(\theta)=|E_t+E_g|^2=|\delta E_i+\beta\alpha E_i\exp
{i\gamma}|^2
\end{equation}
The normalised intensity $I_{g}/I_0$ can then be expressed, with
$\eta_o=\alpha\beta/\delta$,
\begin{equation}
\frac{I_{g}}{I_0}\propto 1 +\eta_o^2+2\eta_o\cos
\gamma\quad\mbox{where}\quad\gamma=k_0l_0+\varphi\label{Eq:output
side interference intensity}
\end{equation}
with $l_0=x_{sg}\sin
 \theta\quad$and $\varphi=k_{\mathrm{surf}}x_{sg}+\varphi_{\mathrm{int}}$.  The relations between $l_0, x_{sg}, \theta$ are shown in
Fig.\,\ref{Fig:InterferenceDiag}.  The frequency and phase of the
interference pattern depend on the slit-groove distance through
the terms $k_{\mathrm{surf}}x_{sg}$, and $k_0x_{sg}\sin\theta$.
The term $\varphi_{\mathrm{int}}$ represents the ``intrinsic"
phase shift due to groove geometry and surface wave generation at
the slit edge.
%%%%%%%%%%%%%%%%%%%%%%%%%%%%%%%%%%%%%%%%%%%%%%%%%%%%%%%%%%%%%%%%%%%%%%%%%%%%%%%%%%%%%%%%%%%%%%%
With the goniometer detector oriented perpendicular to the
structure plane ($\theta=0$), Eq.\,\ref{Eq:output side
interference intensity} simplifies to
\begin{equation}
\frac{I_{g}}{I_0}\propto 1
+\eta_o^2+2\eta_o\cos(k_{\mathrm{surf}}x_{sg}+\varphi_{\mathrm{int}})\label{Eq:fringes}
\end{equation}
Figure \ref{Fig:1stgOutputSidePhase} plots a series of
measurements of the far-field intensity as a function of the
slit-groove distance $x_{sg}$ with the detector angle at
$\theta=0$. The interference term on the right side of
Eq.\,\ref{Eq:fringes} is fit to the data from which
$\varphi_{\mathrm{int}}$ can be determined by extrapolation of
$x_{\mathrm{sg}}$ to zero distance.  The magnitude of the surface
wave propagation vector
$k_{\mathrm{surf}}=2\pi/\lambda_{\mathrm{surf}}$ is also
determined from the fit.  We mesure $\lambda_{\mathrm{surf}}$ to
be $811\pm 8$\,nm in agreement with the input-side experiments
reported earlier\,\cite{GAV06}.  The intrinsic phase
$\varphi_{\mathrm{int}}$ for this groove groove geometry (100\,nm
width and depth) is determined from the plot to be
$\varphi_{\mathrm{int}} = 0.32\,\pi\pm 0.02\,\pi$.
%%%%%%%%%%%%%%%%%%%%%%%%%%%%%%%%%%%%%%%%%%%%%%%%%%%%%%%%%%%%%%%%%%%%%%
%%%%%%%%%%%%%%%%%%%%%%%%%%%%%%%%%%%%%%%%%%%%%%%%%%%%%%%%%%%%%%%%%%%%%%%

\begin{figure}
\centering
\includegraphics[width=0.85\columnwidth]{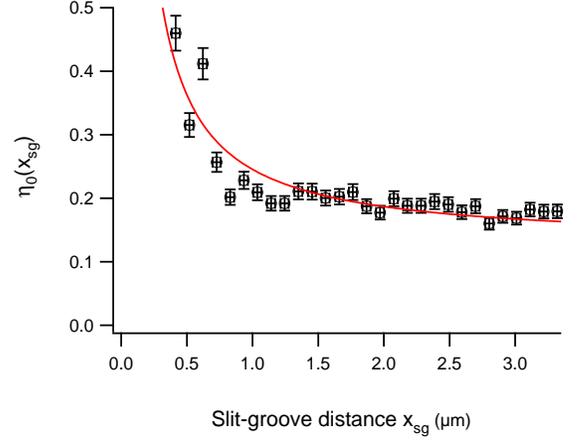}\caption{Plot of $\eta_o$ as a function
of the slit-groove distance $x_{sg}$.  The quantity $\eta_o$ is
fit by Eq.\,\ref{Eq:EtaSlitInputSide} with fitting parameters
$\mu=0.13\pm 0.01$ and $\kappa=0.12\pm 0.01
\mu$m.}\label{Fig:Eta-Thin}
\end{figure}

In addition to the frequency and phase of the interference we have
studied the "visibility" or contrast of the output-side
interference fringes as function of $x_{sg}$. The interference
contrast is defined as
\begin{equation}
C\equiv\frac{I_{max}-I_{min}}{I_{max}+I_{min}}
\end{equation}
where $I_{max},I_{min}$ are adjacent intensity maxima and minima
of the fringes.  According to Eq.\,\ref{Eq:output side
interference intensity} the contrast can be expressed as
\begin{equation}
C=\frac{2\eta_o}{1+\eta_o^2}\quad\mbox{or}\quad\eta_o=\frac{1-\sqrt{1-C^2}}{C}\simeq
\frac{1}{2}C
\end{equation}
Since $\eta_o\propto\alpha\beta=E_g/E_i$, the fractional amplitude
radiating at a groove, a plot of $\eta_o$ as a function of
$x_{sg}$ measures the dependence of this field amplitude (and
therefore the surface wave amplitude) on the slit-groove distance.
Figure \ref{Fig:Eta-Thin} shows a plot of $\eta_o$ as a function
of $x_{sg}$ for narrow-groove structures. The form of the fitted
curve through the data points, an inverse distance dependence with
an additive constant, is given by Eq.\,\ref{Eq:EtaSlitInputSide}
with fitting parameters $\mu,\kappa$ as indicated in the captions
of Fig.\,\ref{Fig:Eta-Thin}.
\begin{equation}
\eta_{o}(x_{sg})=\left(\frac{\kappa}{x_{sg}}+\mu\right)\label{Eq:EtaSlitInputSide}\\
\end{equation}
%%%%%%%%%%%%%%%%%%%%%%%%%%%%%%%%%%%%%%%%%%%%%%%%%%%%%%%%%%%%%%%%%%%%%
%%%%%%%%%%%%%%%%%%%%%%%%%%%%%%%%%%%%%%%%%%%%%%%%%%%%%%%%%%%%%%%%%%%%%%%%

We can now compare these results to predictions of the CDEW model.
Two key predictions are: (1) that the amplitude of the composite
surface wave decreases as the inverse of the distance from the
launch site, and (2) that there is an intrinsic phase delay
$\varphi_{\mathrm{int}}$ of $\pi/2$ between $E_t$, the directly
transmitted wave and $E_{\mathrm{surf}}$ the composite evanescent
wave.  Figure \ref{Fig:Eta-Thin} shows that the contrast (and
therefore the amplitude of $E_g$) fits well a $1/x_{sg}$ behavior
for about the first 3 microns, but then stabilises at a constant
contrast.

\begin{figure}
\includegraphics[width=0.95\columnwidth]{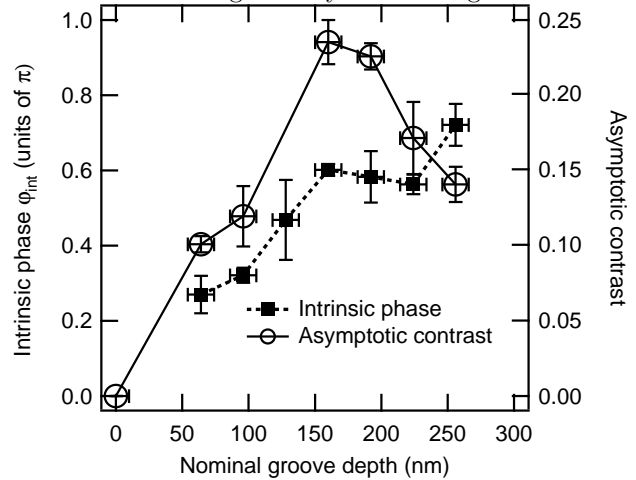}\caption{Intrinsic phase $\varphi_{\mathrm{int}}$ (filled squares,left ordinate)
and contrast (open circles, right ordinate) vs. groove
depth.}\label{Fig:Phase-vs-Depth}
\end{figure}

It appears therefore that there are two components to the surface
wave amplitude: a rapidly decreasing component at short range
followed by a constant component at longer range. Figure
\ref{Fig:1stgOutputSidePhase} shows that $\varphi_{\mathrm{int}}$
extrapolates to $0.32\,\pi \pm 0.02$ (not $\pi/2$) as $x_{sg}$
approaches zero. However, it is well known that grooves exhibit
``organ-pipe" phase shifts and amplitude resonances when the
effective depth is close to an integer number of quarter
wavelengths\,\cite{VLE03}. In order to investigate this
contribution to the intrinsic phase we measured the contrast and
phase as a function of groove depth. The results are shown in
Fig.\,\ref{Fig:Phase-vs-Depth}. The contrast indeed shows a
maximum near 175\,nm groove depth.  Around this resonance the
phase lag from the groove must be about modulo $2\pi$, and
therefore any residual intrinsic phase between $E_t$ and $E_g$
around the groove resonance must be attributed to the phase delay
between the surface wave $E_{\mathrm{surf}}$ and the directly
transmitted wave $E_t$. Figure \ref{Fig:Phase-vs-Depth} indeed
shows that this residual phase is close to $\pi/2$, the signature
phase lag of the CDEW.  Phase and amplitude data from the
wide-groove studies\,\cite{GAV05} are consistent with results
reported here.

We conclude from these phase and amplitude results that the
surface waves exhibit both CDEW-like and SPP-like properties. The
initial decrease in interference contrast, fitting well a $1/x$
behavior, at small slit-groove distances is consistent with a
diffractive surface perturbation at the slit edge.  The
persistance of essentially constant contrast at slit-groove
distances greater than $\sim 2\,\mu$m indicates the presence of an
an SPP-like long-lived mode propagating along the surface.  It
should be noted, however, that the expected wavelength of a pure
SPP mode on a plane silver service is
$\lambda_{\mathrm{SPP}}=844$\,nm\, \cite{Raether88}, but the
measured (Fig.\,\ref{Fig:1stgOutputSidePhase})
$\lambda_{\mathrm{surf}}=811\pm 8$. The reason for the discrepancy
is unclear but this ``output-side" determination is consistent
with the ``input-side" results \cite{GAV06}.  When the
``organ-pipe" phase contributions are taken into account, the
results of Fig.\,\ref{Fig:Phase-vs-Depth} indicate that the the
persistant SPP-like wave conserves the CDEW signature phase lag of
$\pi/2$. The detailed nature of the conversion from a diffracted
surface wave packet to an SPP guided wave has yet to be explained.
\begin{acknowledgments}
Support from the Minist{\`e}re d{\'e}l{\'e}gu{\'e} {\`a} l'Enseignement sup{\'e}rieur et {\`a}
la Recherche under the programme
ACI-``Nanosciences-Nanotechnologies," the R{\'e}gion Midi-Pyr{\'e}n{\'e}es
[SFC/CR 02/22], and FASTNet [HPRN-CT-2002-00304]\,EU Research
Training Network, is gratefully acknowledged as is support from
the Caltech Kavli Nanoscience Institute and from the AFOSR under
Plasmon MURI FA9550-04-1-0434.  Discussions and technical
assistance from P. Lalanne, R. Mathevet, F. Kalkum, G. Derose, A.
Scherer, D. Pacifici, J. Dionne, R. Walters and H. Atwater are
also gratefully acknowledged.
\end{acknowledgments}

\end{document}